\documentclass[prd,aps,preprintnumbers,nofootinbib,superscriptaddress,notitlepage,floatfix,11pt]{revtex4}
\usepackage{stmaryrd}
\usepackage{epsfig}
\usepackage{amsfonts}
\usepackage{amsmath}
\usepackage{graphicx}
\usepackage{color}
\usepackage{amssymb,amsmath,psfrag,slashed,graphicx}
\usepackage[normalem]{ulem}

\def \nn {\nonumber}

\allowdisplaybreaks

\begin{document}
\title{Searching for tetraquark through weak decays of b-baryons}
\affiliation{School of Physics and Microelectronics, Zhengzhou University, Zhengzhou, Henan 450001, China}
\affiliation{School of Physics, China University of Mining and Technology, Xuzhou 221000, China}
\affiliation{INPAC, Key Laboratory for Particle Astrophysics and Cosmology (MOE),
Shanghai Key Laboratory for Particle Physics and Cosmology,
School of Physics and Astronomy, Shanghai Jiao Tong University, Shanghai 200240, China}

\author{Fei Huang~\footnote{fhuang@sjtu.edu.cn}}
\affiliation{INPAC, Key Laboratory for Particle Astrophysics and Cosmology (MOE),
Shanghai Key Laboratory for Particle Physics and Cosmology,
School of Physics and Astronomy, Shanghai Jiao Tong University, Shanghai 200240, China}

\author{Ye Xing~\footnote{Corresponding author. xingye\_guang@cumt.edu.cn}}
\affiliation{School of Physics, China University of Mining and Technology, Xuzhou 221000, China}

\author{Ji Xu~\footnote{Corresponding author. xuji\_phy@zzu.edu.cn}}
\affiliation{School of Physics and Microelectronics, Zhengzhou University, Zhengzhou, Henan 450001, China}

\begin{abstract}
  After the experimental observation of charged and neutral $Z_c(3900)$, it is of prime importance to pursue searches for additional production modes of these exotic states. In this work, we investigate the possibility to study the $Z_c(3900)$ through weak decays of b-baryons at the LHCb. Decay amplitudes for various processes have been parametrized in terms of the SU(3) irreducible nonperturbative amplitudes. A number of relations for the partial decay widths are deduced from these results. The decay widths of $\Lambda_b^0\to \Lambda^0 Z_c^0(3900)$, $\Xi_b^- \to \Sigma^- Z_{c}^0(3900)$ and $\Xi_b^0 \to \Lambda^0 Z_{c}^0(3900)$ have also been calculated from which some relevant partial decay widths of b-baryons could be estimated. The results presented in this paper can be tested experimentally at hadron colliders in the future.
\end{abstract}

\maketitle

\section{Introduction}
\label{sec:Introduction}
In 2013, the BES\uppercase\expandafter{\romannumeral3} Collaboration analyzed the invariant mass spectrum of $\pi^\pm J/\psi$  in the process $e^+ e^- \to \pi^+ \pi^- J/\psi$ at $\sqrt{s}=4.26~\textrm{GeV}$ \cite{BESIII:2013ris}. The result suggests there exists an interesting substructure in the $Y(4260)\to \pi^+ \pi^- J/\psi$ process in the charmonium region, which became renowned as $Z_c^{\pm}(3900)$. This is an important finding since a charged hadron decaying into a charmonium state plus a charged meson must contain at least four quarks. Subsequently, this finding was confirmed by Belle and CLEO-c experiments \cite{Belle:2013yex,Xiao:2013iha}. Soon after the neutral partner $Z_c^{0}(3900)$ has been also observed \cite{BESIII:2015cld}. Such new particles changed dramatically our understanding of exotic states which can not be included in the conventional quark-antiquark and three-quark schemes of standard spectroscopy.

The observed exotic state $Z_c(3900)$ carries an electric charge and couples to charmonium, this makes it a unique platform for sharp tests of QCD. Further studies on the properties of the exotic hadrons would help to understand the formation of exotic hadron states and the character of the strong force. With no doubt, these discoveries have aroused great enthusiasm in theoretical research. Proposed interpretations for $Z_c(3900)$ include hadroquarkonia, hadronic molecules, tetraquark states, and kinematic effects. There were also attempts to describe these newly discovered charmoniumlike states as excitations of ordinary $c\bar c$ charmonium \cite{Wang:2013cya,Guo:2013sya,Li:2013xia,Zhang:2013aoa,Dias:2013xfa,Braaten:2013boa,Wilbring:2013cha,Chen:2013coa,Liu:2013rxa,Wang:2013hga,
Ali:2011ug,Liu:2013vfa,Chen:2013wca,Aceti:2014uea,Wang:2013vex,Chen:2015igx}. All these attempts aim at revealing the internal quark-gluon structure and explaining the masses and decay widths of $Z_c(3900)$, one can find much pretty encouraging results in these papers and state that the whole picture of exotic multiquark states is clearer than when BES\uppercase\expandafter{\romannumeral3} Collaboration firstly discovered $Z_c(3900)$. However, we should stress that the precise structures of the $Z_c(3900)$ and other ``XYZ'' states remain unknown, there is no consensus as to the underlying dynamics which form these states.

\begin{figure}
\includegraphics[width=0.9\columnwidth]{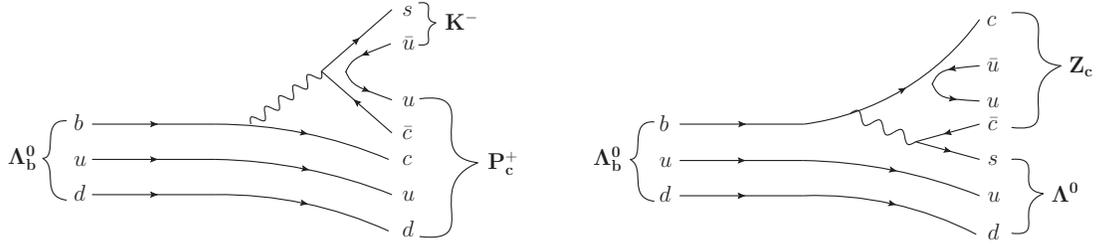}
\caption{Feynman diagrams for illustration which contribute to $\Lambda_b^0 \to K^- P_c$ (left) and the analogous $\Lambda_b^0 \to \Lambda^0 Z_c$ (right).}
\label{Feynman_diagram_for_b_decay}
\end{figure}

Therefore, to further understand the nature of $Z_c(3900)$, it is of prime importance to pursue searches for additional production modes of $Z_c(3900)$ instead of electron-positron collision. There are some proposals such as pion-induced production and $pp$ production of $Z_c(3900)$ \cite{Huang:2015xud,Zhang:2020vfv,D0:2018wyb}. Here, we study the possibility of searching for $Z_c(3900)$ from weak decays of b-baryons at the LHCb. In 2015, the LHCb collaboration has reported two exotic structures $P_c(4380)$ and $P_c(4450)$, firstly observed in the $\Lambda_b^0 \to P_c(\to J/\psi p) K^-$ process \cite{LHCb:2015yax}. From an experimental viewpoint, the decay $\Lambda_b^0 \to \Lambda^0 Z_c $ might be a suitable search channel. As shown in Fig.\ref{Feynman_diagram_for_b_decay}, the tree-level amplitudes of the exclusive decays of $\Lambda_b^0 \to K^- P_c$ and $\Lambda_b^0 \to \Lambda^0 Z_c$ are comparable. Nevertheless, we should stress that constrained by the lack of data and nonperturbative nature, there is no universal factorization approach established to handle production processes of $Z_c$. This gives a barrier for us to predict their production widths through weak decays of b-baryons systematically. On the other hand, the approach of flavor SU(3) symmetry allows us to relate decay modes in the bottom-quark decays in spite of the nonperturbative dynamics of QCD \cite{Savage:1989ub,Gronau:1994rj,He:1998rq,Deshpande:2000jp,Deshpande:1994ii,Chiang:2003pm,
Li:2007bh,Zhou:2015jba,Jiang:2017zwr,Wang:2008rk,Chiang:2008zb,Cheng:2014rfa,He:2014xha,He:2015fwa,He:2015fsa,Lu:2016ogy,Cheng:2016ejf,Cheng:2012xb,Li:2012cfa,Qin:2013tje,
Wang:2017azm,Shi:2017dto,Wang:2018utj,Li:2021rfj,Zhang:2018llc,Huang:2021jxt,Chen:2022asf,He:2006ud}. The diquark model predicts that the charged and neutral $Z_c(3900)$ could be in one octet multiplet of SU(3). Finding the other states in this multiplet will provide crucial evidence for this model.

In this work, we consider nonleptonic decay channels of b-baryons by utilizing flavor SU(3) analysis. Some testable relations for b-baryon decays into a $Z_c$ and a light baryon are presented. These relations can be used as tests for pinning down the suitable model for $Z_c$ states. Very recently, the angular distributions for the decays $\Lambda_b^0\to\Lambda^{0*} J/\psi$, where the $\Lambda^{0*}$ are $\Lambda^0$-type excited states, have been derived in terms of the helicity amplitude technique \cite{Xing:2022uqu,Rui:2022sdc}. We will also use this technique to calculate the partial decay widths of $\Lambda_b^0 \to \Lambda^0 Z_c^0(3900)$, $\Xi_b^- \to \Sigma^- Z_{c}^0(3900)$ and $\Xi_b^0 \to \Lambda^0 Z_{c}^0(3900)$. Having these results at hand, one can get pieces of information on the various related decay channels through flavor SU(3) symmetry. The main motivation of this work is to provide some suggestions which may help experimentalists find new $Z_c$ states or new production modes of already observed $Z_c$ states.

The present paper is arranged as follows. In Sect.~\ref{sec:Production}, we discuss the weak decays of b-baryons whose final states include $Z_c$ and a light baryon. In Sec.~\ref{sec:Analysis}, an analysis of $\Lambda_b^0\to \Lambda^0 Z_c^0(3900)$ will be presented as well as the results of some other decay modes. A discussion based on the results will be also given in this section. Finally, we summarize in the last section.

\section{Production of tetraquark through weak decays of b-baryon}
\label{sec:Production}
In this section, we start with collecting the relevant representations for hadron multiplets under the flavor SU(3) group. The b-baryons contain an antitriplet and a sextet multiplets, they are denoted as $\mathcal{B}$ and $\mathcal{C}$
\begin{eqnarray}
  \left(\mathcal{B}\right)_{i j}=\left(\begin{array}{ccc}
0 & \Lambda_{b}^{0} & \Xi_{b}^{0} \\
-\Lambda_{b}^{0} & 0 & \Xi_{b}^{-} \\
-\Xi_{b}^{0} & -\Xi_{b}^{-} & 0
\end{array}\right), \quad\left(\mathcal{C}\right)_{i j}=\left(\begin{array}{ccc}
\Sigma_{b}^{+} & \frac{\Sigma_{b}^{0}}{\sqrt{2}} & \frac{\Xi_{b}^{\prime 0}}{\sqrt{2}} \\
\frac{\Sigma_{b}^{0}}{\sqrt{2}} & \Sigma_{b}^{-} & \frac{\Xi_{b}^{\prime}}{\sqrt{2}} \\
\frac{\Xi_{b}^{\prime 0}}{\sqrt{2}} & \frac{\Xi_{b}^{\prime-}}{\sqrt{2}} & \Omega_{b}^{-}
\end{array}\right) .
\end{eqnarray}
Light baryons made of three light quarks can group into an SU(3) octet
\begin{eqnarray}
  T_{8}=\left(\begin{array}{ccc}
\frac{1}{\sqrt{2}} \Sigma^{0}+\frac{1}{\sqrt{6}} \Lambda^{0} & \Sigma^{+} & p \\
\Sigma^{-} & -\frac{1}{\sqrt{2}} \Sigma^{0}+\frac{1}{\sqrt{6}} \Lambda^{0} & n \\
\Xi^{-} & \Xi^{0} & -\sqrt{\frac{2}{3}} \Lambda^{0}
\end{array}\right) \,.
\end{eqnarray}
The tetraquark discussed in this work contains at least two light quarks in addition to a $c\bar c$ pair, i.e. [$c\bar c q \bar q$]. Under the flavor SU(3) symmetry, the heavy quark is a singlet, the light quark and light antiquark transform under the flavor SU(3) symmetry as $3\otimes\bar{3}=1+8$. We denote the octet tetraquark component fields as
\begin{eqnarray}
\left(\mathcal{Z}_c\right)_{i}^{j}=\left(\begin{array}{ccc}
\frac{Z_{c\pi^0}}{\sqrt{2}}+\frac{Z_{c\eta_{8}}}{\sqrt{6}} & Z_{c\pi^+} & Z_{cK^{+}} \\
Z_{c\pi^-} & -\frac{Z_{c\pi^0}}{\sqrt{2}}+\frac{Z_{c\eta_{8}}}{\sqrt{6}} & Z_{cK^{0}} \\
Z_{cK^{-}} & Z_{c\overline{K}^{0}} & -\frac{2 Z_{c\eta_{8}}}{\sqrt{6}}
\end{array}\right) \,.
\end{eqnarray}
We will not consider the flavor singlet state to avoid potential octet-singlet mixture complexity.

The weak decay modes of b-baryon into an octet tetraquark and a light octet baryon are discussed in the following. The leading-order effective Hamiltonian is given by
\begin{eqnarray}\label{Hamiltonian}
  \mathcal{H}_{e f f}(b \rightarrow q c \bar{c})=\frac{G_{F}}{\sqrt{2}}\bigg(V_{c b} V_{c q}^{*}\left(C_{1} O_{1}+C_{2} O_{2}\right)\bigg) \,,
\end{eqnarray}
with
\begin{eqnarray}
  O_{1}=\left(\bar{c}_{\alpha} b_{\beta}\right)_{V-A}\left(\bar{q}_{\beta} c_{\alpha}\right)_{V-A}, \quad O_{2}=\left(\bar{c}_{\alpha} b_{\alpha}\right)_{V-A}\left(\bar{q}_{\beta} c_{\beta}\right)_{V-A} \,,
\end{eqnarray}
where $q$ can be $d$ or $s$. The $G_F$ and $V_{ij}$ are Fermi coupling constant and CKM matrix element, respectively. $O_i$ is the low-energy effective operator and $C_i$ is the corresponding Wilson coefficient obtained by integrating out the high energy contributions. In the above, we have neglected contributions from penguin diagrams which are significantly suppressed compared to the tree contributions. The operators $O_i$ transfer under the flavor SU(3) as $\bar 3$, the corresponding quark level transition $b\to c\bar c d/s$ can form a SU(3) triplet with $(H_{3})_{31}=-(H_{3})_{13}=V_{c d}^{*}$ and $(H_{3})_{12}=-(H_{3})_{21}=V_{c s}^{*}$.

For a b-baryon which belongs to the anti-triplet multiplet decays into an octet tetraquark and a light baryon, the corresponding effective Hamiltonian can be constructed as
\begin{eqnarray}\label{FirstH}
	{\cal H}_{\textit{eff}}&=& a_1(\mathcal{B})^{ij}(H_3)_{ij}(\mathcal{Z}_c)^l_k (T_8)^k_l + a_2(\mathcal{B})^{ij}(H_3)_{ik}(\mathcal{Z}_c)^k_l (T_8)^l_j \nn\\
&&+a_3(\mathcal{B})^{ij}(H_3)_{il}(\mathcal{Z}_c)^k_j (T_8)^l_k +a_4(\mathcal{B})^{ij}(H_3)_{kl}(\mathcal{Z}_c)^k_i (T_8)^l_j \,.
\end{eqnarray}

For a b-baryon  belongs to the sextet multiplet, the effective Hamiltonian reads
\begin{eqnarray}\label{SecondH}
	{\cal H}_{\textit{eff}}&=& b_1(\mathcal{C})^{ij}(H_3)_{ik}(\mathcal{Z}_c)^k_l (T_8)^l_j + b_2(\mathcal{C})^{ij}(H_3)_{il}(\mathcal{Z}_c)^k_j (T_8)^l_k +b_3(\mathcal{C})^{ij}(H_3)_{kl}(\mathcal{Z}_c)^k_i (T_8)^l_j \,.
\end{eqnarray}

\begin{figure}
\includegraphics[width=0.9\columnwidth]{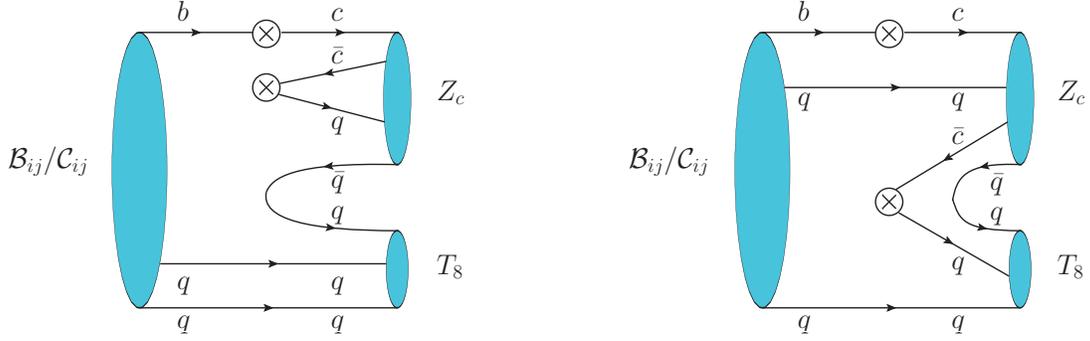}
\caption{Feynman diagrams for a b-baryon decays into an octet tetraquark and a light baryon.}
\label{2.1TbcBM1}
\end{figure}

Where the $a_i$ and $b_i$ are SU(3) irreducible nonperturbative amplitudes. Feynman diagrams for these modes are given in Fig.\,\ref{2.1TbcBM1}. After expanding the above effective Hamiltonian, we can obtain the individual decay amplitudes which are collected in Table.\,\ref{tab:sin} and Table\,\ref{tab:sex}. Many properties concerning weak decays of b-baryons to a tetraquark and a light baryon can be read off from these results. We present some of the interesting properties in the following.

  \begin{table}
\caption{Amplitudes for b-baryon (antitriplet) decays into a tetraquark and a light baryon.}\label{tab:sin}\begin{tabular}{|cc||cc| c c}\hline\hline
channel & amplitude&channel & amplitude \\\hline

$\Lambda_b^0\to   Z_{c\pi^+}  \Sigma^- $ & $ \left(2 a_1+a_2+a_3-a_4\right)\times V_{cs}^*$&
$\Lambda_b^0\to   Z_{c\pi^0}  n $ & $ \frac{\left(a_3-a_4\right) }{\sqrt{2}}\times V_{cd}^*$\\
$\Lambda_b^0\to   Z_{c K^-}  p $ & $ \left(2 a_1+a_2\right) \times V_{cs}^*$&
$\Lambda_b^0\to   Z_{c\pi^-}   p $ & $ \left(a_4-a_3\right) \times V_{cd}^*$\\
$\Lambda_b^0\to   Z_{c\pi^0}  \Sigma^0 $ & $ \left(2 a_1+a_2+a_3-a_4\right) \times V_{cs}^*$&
$\Lambda_b^0\to   Z_{c K^+}  \Sigma^- $ & $ \left(a_4-a_2\right) \times  V_{cd}^*$\\
$\Lambda_b^0\to   Z_{c\pi^-}   \Sigma^+ $ & $ \left(2 a_1+a_2+a_3-a_4\right) \times V_{cs}^*$&
$\Lambda_b^0\to   Z_{c K^0}  \Sigma^0 $ & $ \frac{\left(a_2-a_4\right) }{\sqrt{2}}\times V_{cd}^*$\\
$\Lambda_b^0\to   Z_{c \overline K^0}  n $ & $ \left(2 a_1+a_2\right) \times V_{cs}^*$&
$\Lambda_b^0\to   Z_{c K^0}  \Lambda^0 $ & $ -\frac{\left(a_2-2 a_3+a_4\right) }{\sqrt{6}}\times V_{cd}^*$\\
$\Lambda_b^0\to   Z_{c\eta_8}  \Lambda^0 $ & $ \frac{1}{3} \left(6 a_1+a_2+a_3+a_4\right)\times  V_{cs}^*$&
$\Lambda_b^0\to   Z_{c\eta_8}  n $ & $ \frac{\left(2 a_2-a_3-a_4\right) }{\sqrt{6}}\times V_{cd}^*$\\
$\Xi_b^0\to   Z_{c K^-}  \Sigma^+ $ & $ \left(a_3-a_4\right)\times  V_{cs}^*$&
$\Xi_b^0\to   Z_{c\pi^0}  \Lambda^0 $ & $ -\frac{\left(a_2+a_3-2 a_4\right) }{2 \sqrt{3}}\times V_{cd}^*$\\
$\Xi_b^0\to   Z_{c \overline K^0}  \Sigma^0 $ & $ \frac{\left(a_4-a_3\right) }{\sqrt{2}}\times V_{cs}^*$&
$\Xi_b^0\to   Z_{c\pi^-}   \Sigma^+ $ & $ -\left(2 a_1+a_2\right)\times  V_{cd}^*$\\
$\Xi_b^0\to   Z_{c \overline K^0}  \Lambda^0 $ & $ \frac{\left(-2 a_2+a_3+a_4\right) }{\sqrt{6}}\times V_{cs}^*$&
$\Xi_b^0\to   Z_{c\pi^0}  \Sigma^0 $ & $ -\frac{1}{2} \left(4 a_1+a_2+a_3\right)\times  V_{cd}^*$\\
$\Xi_b^-\to   Z_{c K^-}  \Sigma^0 $ & $ \frac{\left(a_4-a_3\right) }{\sqrt{2}}\times V_{cs}^*$&
$\Xi_b^0\to   Z_{c \overline K^0}  n $ & $ -\left(2 a_1+a_3\right) \times V_{cd}^*$\\
$\Xi_b^-\to   Z_{c K^-}  \Lambda^0 $ & $ \frac{\left(2 a_2-a_3-a_4\right) }{\sqrt{6}}\times V_{cs}^*$&
$\Xi_b^0\to   Z_{c\eta_8}  \Lambda^0 $ & $ -\frac{1}{6} \left(12 a_1+5 a_2+5 a_3-4 a_4\right)\times V_{cd}^*$\\
$\Xi_b^-\to   Z_{c \overline K^0}  \Sigma^- $ & $ \left(a_4-a_3\right)\times  V_{cs}^*$&
$\Xi_b^0\to   Z_{c\eta_8}  \Sigma^0 $ & $ -\frac{\left(a_2+a_3-2 a_4\right) }{2 \sqrt{3}}\times V_{cd}^*$\\
&&
$\Xi_b^0\to   Z_{c K^-}  p $ & $ -\left(2 a_1+a_2+a_3-a_4\right) \times V_{cd}^*$\\
&&
$\Xi_b^0\to   Z_{c\pi^+}  \Sigma^- $ & $ -\left(2 a_1+a_3\right)\times V_{cd}^*$\\
&&
$\Xi_b^-\to   Z_{c K^-}  n $ & $ \left(a_4-a_2\right)\times V_{cd}^*$\\
&&
$\Xi_b^-\to   Z_{c\eta_8}  \Sigma^- $ & $ -\frac{\left(a_2+a_3-2 a_4\right) }{\sqrt{6}}\times V_{cd}^*$\\
&&
$\Xi_b^-\to   Z_{c\pi^0}  \Sigma^- $ & $ \frac{\left(a_3-a_2\right) }{\sqrt{2}}\times V_{cd}^*$\\
&&
$\Xi_b^-\to   Z_{c\pi^-}   \Lambda^0 $ & $ -\frac{\left(a_2+a_3-2 a_4\right) }{\sqrt{6}}\times V_{cd}^*$\\
&&
$\Xi_b^-\to   Z_{c\pi^-}   \Sigma^0 $ & $ \frac{\left(a_2-a_3\right) }{\sqrt{2}}\times V_{cd}^*$\\\hline
\end{tabular}
\end{table}

\begin{table}
\caption{Amplitudes for b-baryon (sextet) decays into a tetraquark and a light baryon.}\label{tab:sex}\begin{tabular}{|cc||cc|c|c|c|c}\hline\hline
channel & amplitude& channel & amplitude  \\\hline
$\Sigma_{b}^{+}\to   Z_{c\pi^+}  \Lambda^0 $ & $ \frac{\left(b_1+b_2-b_3\right) }{\sqrt{6}}\times V_{cs}^*$&
$\Sigma_{b}^{+}\to   Z_{c\pi^0}  p $ & $ -\frac{\left(b_2+b_3\right) }{\sqrt{2}}\times V_{cd}^*$\\
$\Sigma_{b}^{+}\to   Z_{c\pi^+}  \Sigma^0 $ & $ \frac{\left(b_1-b_2-b_3\right) }{\sqrt{2}}\times V_{cs}^*$&
$\Sigma_{b}^{+}\to   Z_{c K^+}  \Sigma^0 $ & $ \frac{\left(b_3-b_1\right) }{\sqrt{2}}\times V_{cd}^*$\\
$\Sigma_{b}^{+}\to   Z_{c\pi^0}  \Sigma^+ $ & $ \frac{\left(-b_1+b_2+b_3\right) }{\sqrt{2}}\times V_{cs}^*$&
$\Sigma_{b}^{+}\to   Z_{c K^+}  \Lambda^0 $ & $ \frac{\left(-b_1+2 b_2+b_3\right) }{\sqrt{6}}\times V_{cd}^*$\\
$\Sigma_{b}^{+}\to   Z_{c\eta_8}  \Sigma^+ $ & $ \frac{\left(b_1+b_2+b_3\right) }{\sqrt{6}}\times V_{cs}^*$&
$\Sigma_{b}^{+}\to   Z_{c K^0}  \Sigma^+ $ & $ -b_1 \times V_{cd}^*$\\
$\Sigma_{b}^{+}\to   Z_{c \overline K^0}  p $ & $ b_1 \times V_{cs}^*$&
$\Sigma_{b}^{+}\to   Z_{c\eta_8}  p $ & $ \frac{\left(2 b_1-b_2-b_3\right) }{\sqrt{6}}\times V_{cd}^*$\\
$\Sigma_{b}^{0}\to   Z_{c\pi^+}  \Sigma^- $ & $ \frac{\left(b_1-b_2-b_3\right) }{\sqrt{2}}\times V_{cs}^*$&
$\Sigma_{b}^{+}\to   Z_{c\pi^+}  n $ & $ -b_2 \times V_{cd}^*$\\
$\Sigma_{b}^{0}\to   Z_{c\pi^0}  \Lambda^0 $ & $ -\frac{\left(b_1+b_2-b_3\right) }{\sqrt{6}}\times V_{cs}^*$&
$\Sigma_{b}^{0}\to   Z_{c\pi^0}  n $ & $ \frac{1}{2} \left(b_2-b_3\right)\times  V_{cd}^*$\\
$\Sigma_{b}^{0}\to   Z_{c\pi^-}   \Sigma^+ $ & $ \frac{\left(-b_1+b_2+b_3\right) }{\sqrt{2}}\times V_{cs}^*$&
$\Sigma_{b}^{0}\to   Z_{c\pi^-}   p $ & $ -\frac{\left(b_2+b_3\right) }{\sqrt{2}}\times V_{cd}^*$\\
$\Sigma_{b}^{0}\to   Z_{c \overline K^0}  n $ & $ \frac{b_1 }{\sqrt{2}}\times V_{cs}^*$&
$\Sigma_{b}^{0}\to   Z_{c K^0}  \Lambda^0 $ & $ \frac{\left(-b_1+2 b_2+b_3\right) }{2 \sqrt{3}}\times V_{cd}^*$\\
$\Sigma_{b}^{0}\to   Z_{c\eta_8}  \Sigma^0 $ & $ -\frac{\left(b_1+b_2+b_3\right) }{\sqrt{6}}\times V_{cs}^*$&
$\Sigma_{b}^{0}\to   Z_{c K^0}  \Sigma^0 $ & $ \frac{1}{2} \left(b_1+b_3\right) \times V_{cd}^*$\\
$\Sigma_{b}^{0}\to   Z_{c K^-}  p $ & $ -\frac{b_1 }{\sqrt{2}}\times V_{cs}^*$&
$\Sigma_{b}^{0}\to   Z_{c\eta_8}  n $ & $ \frac{\left(2 b_1-b_2-b_3\right) }{2 \sqrt{3}}\times V_{cd}^*$\\
$\Sigma_{b}^{-}\to   Z_{c\pi^0}  \Sigma^- $ & $ \frac{\left(-b_1+b_2+b_3\right) }{\sqrt{2}}\times V_{cs}^*$&
$\Sigma_{b}^{0}\to   Z_{c K^+}  \Sigma^- $ & $ \frac{\left(b_3-b_1\right) }{\sqrt{2}}\times V_{cd}^*$\\
$\Sigma_{b}^{-}\to   Z_{c\pi^-}   \Lambda^0 $ & $ -\frac{\left(b_1+b_2-b_3\right) }{\sqrt{6}}\times V_{cs}^*$&
$\Sigma_{b}^{-}\to   Z_{c\pi^-}   n $ & $ -b_3 \times V_{cd}^*$\\
$\Sigma_{b}^{-}\to   Z_{c\pi^-}   \Sigma^0 $ & $ \frac{\left(b_1-b_2-b_3\right) }{\sqrt{2}}\times V_{cs}^*$&
$\Sigma_{b}^{-}\to   Z_{c K^0}  \Sigma^- $ & $ b_3\times  V_{cd}^*$\\
$\Sigma_{b}^{-}\to   Z_{c K^-}  n $ & $ -b_1 \times V_{cs}^*$&
$\Xi_{b}^{\prime0}\to   Z_{c\pi^0}  \Sigma^0 $ & $ \frac{\left(b_1+b_2\right) }{2 \sqrt{2}}\times V_{cd}^*$\\
$\Sigma_{b}^{-}\to   Z_{c\eta_8}  \Sigma^- $ & $ -\frac{\left(b_1+b_2+b_3\right) }{\sqrt{6}}\times V_{cs}^*$&
$\Xi_{b}^{\prime0}\to   Z_{c\pi^0}  \Lambda^0 $ & $ \frac{\left(b_1+b_2+2 b_3\right) }{2 \sqrt{6}}\times V_{cd}^*$\\
$\Xi_{b}^{\prime0}\to   Z_{c \overline K^0}  \Sigma^0 $ & $ -\frac{1}{2} \left(b_2+b_3\right) \times V_{cs}^*$&
$\Xi_{b}^{\prime0}\to   Z_{c\pi^-}   \Sigma^+ $ & $ \frac{b_1 }{\sqrt{2}}\times V_{cd}^*$\\
$\Xi_{b}^{\prime0}\to   Z_{c \overline K^0}  \Lambda^0 $ & $ -\frac{\left(2 b_1-b_2+b_3\right) }{2 \sqrt{3}}\times V_{cs}^*$&
$\Xi_{b}^{\prime0}\to   Z_{c \overline K^0}  n $ & $ -\frac{b_2 }{\sqrt{2}}\times V_{cd}^*$\\
$\Xi_{b}^{\prime0}\to   Z_{c K^-}  \Sigma^+ $ & $ \frac{\left(b_2+b_3\right) }{\sqrt{2}}\times V_{cs}^*$&
$\Xi_{b}^{\prime0}\to   Z_{c K^-}  p $ & $ \frac{\left(b_1-b_2-b_3\right) }{\sqrt{2}}\times V_{cd}^*$\\
$\Xi_{b}^{\prime-}\to   Z_{c K^-}  \Lambda^0 $ & $ \frac{\left(2 b_1-b_2+b_3\right) }{2 \sqrt{3}}\times V_{cs}^*$&
$\Xi_{b}^{\prime0}\to   Z_{c\eta_8}  \Lambda^0 $ & $ -\frac{\left(b_1+b_2\right) }{2 \sqrt{2}}\times V_{cd}^*$\\
$\Xi_{b}^{\prime-}\to   Z_{c \overline K^0}  \Sigma^- $ & $ -\frac{\left(b_2+b_3\right) }{\sqrt{2}}\times V_{cs}^*$&
$\Xi_{b}^{\prime0}\to   Z_{c\eta_8}  \Sigma^0 $ & $ \frac{\left(b_1+b_2-2 b_3\right) }{2 \sqrt{6}}\times V_{cd}^*$\\
$\Xi_{b}^{\prime-}\to   Z_{c K^-}  \Sigma^0 $ & $ -\frac{1}{2} \left(b_2+b_3\right) \times V_{cs}^*$&
$\Xi_{b}^{\prime0}\to   Z_{c\pi^+}  \Sigma^- $ & $ \frac{b_2 }{\sqrt{2}}\times V_{cd}^*$\\
&&
$\Xi_{b}^{\prime-}\to   Z_{c\pi^-}   \Sigma^0 $ & $ \frac{1}{2} \left(b_2-b_1\right) \times V_{cd}^*$\\
&&
$\Xi_{b}^{\prime-}\to   Z_{c K^-}  n $ & $ \frac{\left(b_1-b_3\right) }{\sqrt{2}}\times V_{cd}^*$\\
&&
$\Xi_{b}^{\prime-}\to   Z_{c\pi^0}  \Sigma^- $ & $ \frac{1}{2} \left(b_1-b_2\right) \times V_{cd}^*$\\
&&
$\Xi_{b}^{\prime-}\to   Z_{c\eta_8}  \Sigma^- $ & $ \frac{\left(b_1+b_2-2 b_3\right) }{2 \sqrt{3}}\times V_{cd}^*$\\
&&
$\Xi_{b}^{\prime-}\to   Z_{c\pi^-}   \Lambda^0 $ & $ \frac{\left(b_1+b_2+2 b_3\right) }{2 \sqrt{3}}\times V_{cd}^*$\\
&&
$\Omega_{b}^{-}\to   Z_{c K^-}  \Lambda^0 $ & $ \frac{\left(-2 b_1+b_2+2 b_3\right) }{\sqrt{6}}\times V_{cd}^*$\\
&&
$\Omega_{b}^{-}\to   Z_{c \overline K^0}  \Sigma^- $ & $ b_2 \times V_{cd}^*$\\
&&
$\Omega_{b}^{-}\to   Z_{c K^-}  \Sigma^0 $ & $ \frac{b_2 }{\sqrt{2}}\times V_{cd}^*$\\\hline
\end{tabular}
\end{table}

\begin{enumerate}
  \item Table~\ref{tab:sin} and Table~\ref{tab:sex} are arranged according to the decay amplitude's dependence on CKM matrix elements, $c\to s$ transition is proportional to $|V_{cs}^*|\sim 1$, while $c\to d$ transition has a Cabibbo suppressed CKM matrix element $|V_{cd}^*|\sim 0.2$.
  \item A number of relations for different decay widths can be readily deduced from Table.\,\ref{tab:sin}:
\begin{eqnarray*}
    &&\Gamma(\Xi_b^-\to Z_{c\pi^0} \Sigma^-) = \Gamma(\Xi_b^-\to Z_{c\pi^-}  \Sigma^0) \,,\\
    &&\Gamma(\Lambda_b^0\to Z_{c \overline K^0} n)= \Gamma(\Lambda_b^0\to Z_{c K^-} p) \,,\qquad \Gamma(\Lambda_b^0\to Z_{c\pi^0} n) = \frac{1}{2}\Gamma(\Lambda_b^0\to Z_{c\pi^-}  p) \,,\\
    &&\Gamma(\Xi_b^0\to Z_{c\pi^+} \Sigma^-)= \Gamma(\Xi_b^0\to Z_{c \overline K^0} n) \,,\qquad \Gamma(\Xi_b^0\to Z_{c \overline K^0} \Lambda^0)= \Gamma(\Xi_b^-\to Z_{c K^-} \Lambda^0) \,,\\
&&\Gamma(\Lambda_b^0\to Z_{c\pi^+} \Sigma^-)= \Gamma(\Lambda_b^0\to Z_{c\pi^0} \Sigma^0) = \Gamma(\Lambda_b^0\to Z_{c\pi^-}  \Sigma^+) \,,\\
    &&\Gamma(\Lambda_b^0\to Z_{c K^+} \Sigma^-)= 2\Gamma(\Lambda_b^0\to Z_{c K^0} \Sigma^0)= \Gamma(\Xi_b^-\to Z_{c K^-} n) \,,\\
    &&\Gamma(\Xi_b^0\to Z_{c\pi^0} \Lambda^0)= \Gamma(\Xi_b^0\to Z_{c\eta_8} \Sigma^0)= \frac{1}{2}\Gamma(\Xi_b^-\to Z_{c\pi^-}  \Lambda^0)= \frac{1}{2}\Gamma(\Xi_b^-\to Z_{c\eta_8} \Sigma^-) \,,\\
    &&\Gamma(\Xi_b^0\to Z_{c \overline K^0} \Sigma^0)= \frac{1}{2}\Gamma(\Xi_b^0\to Z_{c K^-} \Sigma^+)= \frac{1}{2}\Gamma(\Xi_b^-\to Z_{c \overline K^0} \Sigma^-)= \Gamma(\Xi_b^-\to Z_{c K^-} \Sigma^0) \,.
\end{eqnarray*}
And the relations deduced from Table.\,\ref{tab:sex}:
\begin{eqnarray*}
    &&\Gamma(\Xi_{b}^{\prime0}\to Z_{c \overline K^0} \Lambda^0)= \Gamma(\Xi_{b}^{\prime-}\to Z_{c K^-} \Lambda^0) \,,\\
    &&\Gamma(\Xi_{b}^{\prime0}\to Z_{c\eta_8} \Sigma^0)= \frac{1}{2}\Gamma(\Xi_{b}^{\prime-}\to Z_{c\eta_8} \Sigma^-) \,, \qquad \Gamma(\Xi_{b}^{\prime-}\to Z_{c\pi^0} \Sigma^-)= \Gamma(\Xi_{b}^{\prime-}\to Z_{c\pi^-}  \Sigma^0) \,,\\
&&\Gamma(\Sigma_{b}^{+}\to Z_{c\pi^0} p)= \Gamma(\Sigma_{b}^{0}\to Z_{c\pi^-}  p) \,, \qquad \Gamma(\Sigma_{b}^{+}\to Z_{c K^+} \Lambda^0) = 2\Gamma(\Sigma_{b}^{0}\to Z_{c K^0} \Lambda^0) \,,\\
    &&\Gamma(\Sigma_{b}^{+}\to Z_{c\eta_8} p)= 2\Gamma(\Sigma_{b}^{0}\to Z_{c\eta_8} n) \,, \qquad   \Gamma(\Sigma_{b}^{-}\to Z_{c\pi^-}  n)= \Gamma(\Sigma_{b}^{-}\to Z_{c K^0} \Sigma^-) \,,\\
    &&\Gamma(\Xi_{b}^{\prime0}\to Z_{c\pi^0} \Lambda^0)= \frac{1}{2}\Gamma(\Xi_{b}^{\prime-}\to Z_{c\pi^-}  \Lambda^0) \,, \qquad \Gamma(\Xi_{b}^{\prime0}\to Z_{c\pi^0} \Sigma^0)= \Gamma(\Xi_{b}^{\prime0}\to Z_{c\eta_8} \Lambda^0) \,,\\
&&\Gamma(\Sigma_{b}^{+}\to Z_{c\pi^+} \Lambda^0)=  \Gamma(\Sigma_{b}^{0}\to Z_{c\pi^0} \Lambda^0)= \Gamma(\Sigma_{b}^{-}\to Z_{c\pi^-}  \Lambda^0) \,,\\
    &&\Gamma(\Sigma_{b}^{+}\to Z_{c K^+} \Sigma^0)= \Gamma(\Sigma_{b}^{0}\to Z_{c K^+} \Sigma^-)= \Gamma(\Xi_{b}^{\prime-}\to Z_{c K^-} n) \,,\\
    &&\Gamma(\Sigma_{b}^{+}\to Z_{c\eta_8} \Sigma^+)= \Gamma(\Sigma_{b}^{0}\to Z_{c\eta_8} \Sigma^0)= \Gamma(\Sigma_{b}^{-}\to Z_{c\eta_8} \Sigma^-) \,,\\
    &&\Gamma(\Sigma_{b}^{+}\to Z_{c \overline K^0} p)= 2\Gamma(\Sigma_{b}^{0}\to Z_{c \overline K^0} n)= 2\Gamma(\Sigma_{b}^{0}\to Z_{c K^-} p)= \Gamma(\Sigma_{b}^{-}\to Z_{c K^-} n) \,,\\
    &&\Gamma(\Xi_{b}^{\prime0}\to Z_{c \overline K^0} \Sigma^0)= \frac{1}{2}\Gamma(\Xi_{b}^{\prime0}\to Z_{c K^-} \Sigma^+)= \frac{1}{2}\Gamma(\Xi_{b}^{\prime-}\to Z_{c \overline K^0} \Sigma^-)= \Gamma(\Xi_{b}^{\prime-}\to Z_{c K^-} \Sigma^0) \,,\\
    &&\Gamma(\Sigma_{b}^{+}\to Z_{c\pi^+} n)= 2\Gamma(\Xi_{b}^{\prime0}\to Z_{c\pi^+} \Sigma^-)= 2\Gamma(\Xi_{b}^{\prime0}\to Z_{c \overline K^0} n)= \Gamma(\Omega_{b}^{-}\to Z_{c \overline K^0} \Sigma^-) \\
    &&= 2\Gamma(\Omega_{b}^{-}\to Z_{c K^-} \Sigma^0) \,,\\
    &&\Gamma(\Sigma_{b}^{+}\to Z_{c\pi^+} \Sigma^0)= \Gamma(\Sigma_{b}^{+}\to Z_{c\pi^0} \Sigma^+)= \Gamma(\Sigma_{b}^{0}\to Z_{c\pi^+} \Sigma^-)= \Gamma(\Sigma_{b}^{0}\to Z_{c\pi^-}  \Sigma^+)\\ &&=\Gamma(\Sigma_{b}^{-}\to Z_{c\pi^0} \Sigma^-)= \Gamma(\Sigma_{b}^{-}\to Z_{c\pi^-}  \Sigma^0) \,.
\end{eqnarray*}

  \item Some theoretical researches predict that the charged and neutral $Z_c(3900)$ belong to the same octet multiplet of SU(3) group. Based on the different valence quark components, we take $Z_c^{\pm}(3900)$ as $Z_{c\pi^{\pm}}$ and $Z_c^{0}(3900)$ as $Z_{c\pi^{0}}$ in this work. Therefore, once a few branching fractions have been measured or calculated in the future, some of these relations shown above may provide hints for the exploration of new decay modes. The process $\Lambda_b^0 \to \Lambda^0 \, Z_c^0(3900)$ would be analyzed for instance in the next section.
\end{enumerate}

It is necessary to point out that the above relations between different decay widths are obtained in the flavor SU(3) symmetry limit, where the mass differences between final state hadrons have been ignored. Besides, the hadronization processes whose information contained in different decay constants and form factors would also affect the relations derived in this paper. It is widely believed that the symmetry breaking in
bottom-quark decays is pretty small in comparison with charm-quark decays, hence we expect that the relations derived in our analysis basically hold and many of them can be precisely examined in the future.

\section{Analysis of $\Lambda_b^0 \to \Lambda^0 \, Z_c^0(3900)$ by helicity amplitude technique}
\label{sec:Analysis}
We carry out an analysis of the $\Lambda_b^0$ decay $\Lambda_b^0 \to \Lambda^0 \, Z_c^0(3900)$ in terms of the helicity amplitude technique. As we will see later the helicity amplitudes are particularly convenient for expressing
various observable quantities in the heavy quark decays. The amplitude $\mathcal{M}(\Lambda_b^0 \to \Lambda^0 \, Z_c^0(3900))$ is induced by the $b\to s c \bar c$ transition whose effective Hamiltonian has been presented in Eq.\,(\ref{Hamiltonian}). The factorizable diagram of this process is enhanced by color factor compared with the non-factorizable one. One may adopt the factorization ansatz to predict its decay width, the amplitude is
\begin{eqnarray}\label{factorization}
\mathcal{M}\left(\Lambda_{b}^0 \rightarrow \Lambda^0 \, Z_c^0(3900) \right) &=& \frac{G_{F}}{\sqrt{2}} V_{c b} V_{c s}^{*} a_{2} f_{Z_c} M_{Z_c}\left\langle\Lambda^{0}\left|(\bar{s} b)_{V-A}^{\mu}\right| \Lambda_{b}^0 \right\rangle \epsilon_{\mu}^{*}(s_{Z_c}) \,,
\end{eqnarray}
where $a_{2} = C_{1}+C_{2} / N_{c}$, with the Wilson coefficients being $C_1=-0.248$ and $C_2=1.107$ \cite{Buchalla:1995vs,Wang:2008sm}. $f_{Z_c}$ and $M_{Z_c}$ are the decay constant and mass of $Z_c^0(3900)$ respectively, they are important spectroscopic parameters of an exotic multiquark state, which have been evaluated to be $f_{Z_c}=0.0051~\textrm{GeV}$ and $M_{Z_c}=3.901~\textrm{GeV}$ \cite{Wang:2013vex,Agaev:2017tzv}.

The hadron matrix element $\left\langle\Lambda^{0}\left|(\bar{s} b)_{V-A}^{\mu}\right| \Lambda_{b}^0\right\rangle$ in Eq.\,(\ref{factorization}) can be parameterized by the weak transition form factors \cite{Mott:2011cx,Wang:2015ndk,Wang:2009hra}
\begin{eqnarray}
\langle \Lambda^{0}(p^\prime, s^\prime)|\bar{s}\gamma^\mu b|\Lambda_b^0(p, s)\rangle&=&\bar{u}(p^\prime, s^\prime)\big(\gamma_\mu f_1+\frac{p_{\Lambda_b^0}^\mu}{m_{\Lambda_b^0}}f_2+\frac{p_{\Lambda^{0}}^\mu}{m_{\Lambda^{0}}}f_3\big)u(p, s) \,,\notag\\
\langle \Lambda^{0}(p^\prime, s^\prime)|\bar{s}\gamma^\mu\gamma_5 b|\Lambda_b^0(p, s)\rangle&=&\bar{u}(p^\prime, s^\prime)\big(\gamma_\mu g_1+\frac{p_{\Lambda_b^0}^\mu}{m_{\Lambda_b^0}}g_2+\frac{p_{\Lambda^{0}}^\mu}{m_{\Lambda^{0}}}g_3\big)\gamma_5u(p, s) \,.\label{12}
\end{eqnarray}

The helicity amplitudes for $\Lambda_{b}^0 \to \Lambda^0$ are given as
 \begin{align}
H_{1 V}\left(s_{\Lambda_{b}^0}=-\frac{1}{2}, s_{\Lambda}=\frac{1}{2}, s_W=1\right) &=H_{1 V}\left(s_{\Lambda_{b}^0}=\frac{1}{2}, s_{\Lambda}=-\frac{1}{2}, s_W=-1\right) \nonumber\\
&=\sqrt{2 s_{-}}\left[f_{1}(q^{2})-\frac{m_{\Lambda_{b}^0}+m_{\Lambda}}{m_{\Lambda_{b}^0}} f_{2}(q^{2})\right], \\
H_{1 V}\left(s_{\Lambda_{b}^0}=\frac{1}{2}, s_{\Lambda}=\frac{1}{2}, s_W=0\right) &=H_{1 V}\left(s_{\Lambda_{b}^0}=-\frac{1}{2}, s_{\Lambda}=-\frac{1}{2}, s_W=0\right) \nonumber\\
&=\sqrt{\frac{s_{-}}{q^{2}}}\left[\left(m_{\Lambda_{b}^0}+m_{\Lambda}\right) f_{1}(q^{2})-\frac{q^{2}}{m_{\Lambda_{b}^0}} f_{2}(q^{2})\right], \\
H_{1 A}\left(s_{\Lambda_{b}^0}=-\frac{1}{2}, s_{\Lambda}=\frac{1}{2}, s_W=1\right) &=-H_{1 A}\left(s_{\Lambda_{b}^0}=\frac{1}{2}, s_{\Lambda}=-\frac{1}{2}, s_W=-1\right) \nonumber\\
&=\sqrt{2 s_{+}}\left[g_{1}(q^{2})+\frac{m_{\Lambda_{b}^0}-m_{\Lambda}}{m_{\Lambda_{b}^0}} g_{2}(q^{2})\right], \\
H_{1 A}\left(s_{\Lambda_{b}^0}=\frac{1}{2}, s_{\Lambda}=\frac{1}{2}, s_W=0\right) &=-H_{1 A}\left(s_{\Lambda_{b}^0}=-\frac{1}{2}, s_{\Lambda}=-\frac{1}{2}, s_W=0\right) \nonumber\\
&=\sqrt{\frac{s_{+}}{q^{2}}}\left[\left(m_{\Lambda_{b}^0}-m_{\Lambda}\right) g_{1}(q^{2})+\frac{q^{2}}{m_{\Lambda_{b}^0}} g_{2}(q^{2})\right].
\end{align}
Total helicity amplitude can be written as $H_{1}=H_{1 V}-H_{1A}$. The form factors in helicity amplitude are studied in the full quark model wave function (MCN) which can be expressed as \cite{Mott:2011cx}
\begin{eqnarray}
f(q^2)&=&(a_0+a_2 p_{\Lambda}^2+a_4 p_{\Lambda}^4)\exp\left(-\frac{6m_q^2 p_{\Lambda}^2}{2\tilde{m}_\Lambda^2(\alpha_{\Lambda_b^0}^2+\alpha_{\Lambda}^2)}\right) \,,\nonumber\\
p_{\Lambda}&=&\frac{m_{\Lambda_{b}^0}}{2}\left[\left(1-\frac{m_{\Lambda}^{2}}{m_{\Lambda_{b}^0}^2}\right)^{2}
-2\left(1+\frac{m_{\Lambda}^{2}}{m_{\Lambda_{b}^0}^2}\right)\frac{q^{2}}{m_{\Lambda_{b}^0}^{2}}
+\left(\frac{q^{2}}{m_{\Lambda_{b}^0}^{2}}\right)^2\right] \,. \label{lam}
\end{eqnarray}
with $\tilde{m}_\Lambda=m_{u}+m_{d}+m_{s}$. Different $f_i$ corresponds to different $a_{0}, a_{2}, a_{4}$ whose values are given in Table~\ref{formfactors} in the MCN model.

\begin{table}[htbp!]
\caption{Input parameters for spin-$1/2$ baryon $\Lambda^{0}$ in MCN quark model. }\label{formfactors}\begin{tabular}{|c|c|c|c|c|c|c|c|}\hline\hline
\multicolumn{4}{|c|}{ $\Lambda^{0}$}\\\hline \hline
form factor&$a_0$&$a_2$&$a_4$\\\hline
$f^{+}_1$&1.21&0.319&-0.0177\\\hline
$f^{+}_2$&-0.202&-0.219&-0.0103\\\hline
$f^{+}_3$&-0.0615&0.0102&-0.00139\\\hline
$g^{+}_1$&0.927&0.104&-0.00553\\\hline
$g^{+}_2$&-0.236&-0.233&0.011\\\hline
$g^{+}_3$&0.0756&0.0195&-0.00115\\\hline
\multicolumn{2}{|c|}{$\alpha_{\Lambda^{0}}=0.387$}&\multicolumn{2}{|c|}{$\alpha_{\Lambda_{b}^0}=0.443$}\\\hline
\end{tabular}
\end{table}
After the two-body phase space integration, the decay width for $\Lambda_b^0\to \Lambda^0 Z_{c}^0(3900)$ is then formulated as
\begin{eqnarray}
\Gamma(\Lambda_b^0\to \Lambda^0 Z_{c}^0(3900))&=&\sum_{s_{\Lambda_{b}^0},s_{\Lambda}}\frac{ |\vec{p}_{\Lambda}|}{8\pi m^2_{\Lambda_b^0}}\frac{1}{2}|\mathcal{M}(\Lambda_b^0 \to \Lambda^0 Z_{c}^0(3900))|^2 \,\nonumber\\
&=&\sum_{s_{\Lambda_{b}^0},s_{\Lambda}}\frac{ |\vec{p}_{\Lambda}|}{8\pi m^2_{\Lambda_b^0}}\left| \frac{G_{F}}{\sqrt{2}} V_{c b} V_{c s}^{*} a_{2} f_{Z_c} M_{Z_c} \sum_{s_{\Lambda}}\sum_{s_{\Lambda}}H_1(s_{\Lambda_{b}^0},s_{\Lambda},s_W)\right|^2 \,.
\end{eqnarray}
Using the form factors in Table.\,\ref{formfactors}, we obtain an order-of-magnitude estimation of partial decay width and branching fraction
\begin{eqnarray}\label{decay_widths1}
\Gamma(\Lambda_b^0\to \Lambda^0 Z_{c}^0(3900)) &=& 8.61\times 10^{-20}~\textrm{GeV} \,, \nn\\
\mathcal{B}(\Lambda_b^0\to \Lambda^0 Z_{c}^0(3900)) &=& 1.93\times 10^{-7} \,.
\end{eqnarray}
The estimated branching fraction of $\Lambda_b^0\to \Lambda^0 Z_{c}^0(3900)$ is at the order $10^{-7}$. According to the experimental result from BES\uppercase\expandafter{\romannumeral3}, $Z_{c}^0(3900)$ would subsequently decay into $\pi^0 J/\psi$ and most of $\Lambda^0$ decay into $p\,\pi^-$. Therefore the cascade decay would be $\Lambda_b^0\to \Lambda^0 Z_{c}^0(3900)\to p \, J/\psi \, \pi^- \pi^0$. It is worth noting that our result is consistent with the transition matrix element $\Lambda_{b}^0 \to \Lambda^{0}$ given in \cite{Hsiao:2015cda}. The matrix elements of $\Xi_b^- \to \Sigma^-$ and $\Xi_b^0 \to \Lambda^0$ have also been calculated in \cite{Hsiao:2015cda} which could be incorporated into the theoretical formalism shown in this section. Similarly, one obtains the estimation of partial decay widths and branching fractions
\begin{eqnarray}\label{decay_widths2}
\Gamma(\Xi_b^- \to \Sigma^- Z_{c}^0(3900)) &=& 8.40\times 10^{-21}~\textrm{GeV} \,, \qquad \Gamma(\Xi_b^0 \to \Lambda^0 Z_{c}^0(3900)) = 2.64\times 10^{-21}~\textrm{GeV} \,,\nn\\
\mathcal{B}(\Xi_b^- \to \Sigma^- Z_{c}^0(3900)) &=& 2.01\times 10^{-8} \,, \qquad\qquad~~  \mathcal{B}(\Xi_b^0 \to \Lambda^0 Z_{c}^0(3900)) = 5.94\times 10^{-9} \,.
\end{eqnarray}
The results displayed above is an order of magnitude lower than Eq.\,(\ref{decay_widths1}) mainly due to the ratio of CKM matrix elements $|V_{cs}^*/V_{cd}^*|^2$. Notably, at this stage the SU(3) symmetry is helpful for figuring out the promising decay channels toward the discovery of $Z_c$ states through b-baryon decay. According to the SU(3) symmetry induced results in Table.\,\ref{tab:sin} and the branching fractions in Eq.\,(\ref{decay_widths1}) and Eq.\,(\ref{decay_widths2}), we collect several channels of b-baryon decay with estimated branching fractions in Table\,\ref{tab:collect}. The presented branching fractions are on the order of $10^{-7}$ or less, at present, LHCb could detect the branching fraction down to $10^{-6}$ among $\Lambda_b^0$ decay modes whose final states contain $J/\psi$ or its excited states plus a proton and light mesons \cite{Workman:2022ynf}. Thus, the decay channels shown in Table\,\ref{tab:collect} can be only observed with a large amount of data in the future, such as the high luminosity LHC.

\begin{table}
\caption{Estimation of branching fractions of b-baryon decay in which the tetraquark appears in the final states.}\label{tab:collect}\begin{tabular}{cccc}\hline\hline
Channel & Branching fraction &  Channel & Branching fraction \\\hline
$\Xi_b^- \to \Sigma^- Z_{c}^0(3900) $ & $ 2.01\times 10^{-8} $  &
$\Xi_b^- \to \Sigma^0 Z_{c}^-(3900) $ & $ 2.01\times 10^{-8} $ \\
$\Xi_b^- \to \Lambda^0 Z_{c}^-(3900) $ & $ 1.26\times 10^{-8} $  &
$\Xi_b^- \to \Sigma^- Z_{c\eta_8} $ & $ 1.26\times 10^{-8} $ \\
$\Xi_b^0 \to \Lambda^0 Z_{c}^0(3900) $ & $ 5.94\times 10^{-9} $  &
$\Xi_b^0 \to \Sigma^0 Z_{c\eta_8} $ & $ 5.94\times 10^{-9} $ \\
$\Lambda_b^0\to \Lambda^0 Z_{c}^0(3900) $ & $ 1.93\times 10^{-7} $  & & \\
\hline\hline
\end{tabular}
\end{table}

It should be stressed here that the results listed in Table\,\ref{tab:collect} are preferred to be regarded as rough estimations rather than accurate ones since they are based on the naive factorization ansatz. To rigorously study the dynamics of nonleptonic two-body b-baryon decays whose final states contain a tetraquark, some popular theoretical approaches such as perturbative QCD (pQCD) \cite{Lu:2000em,Chou:2001bn},  QCD factorization (QCDF) \cite{Beneke:1999br} and SCET would be necessary \cite{Bauer:2000yr,Bauer:2001yt}. This issue goes surely beyond the scope of this paper and is studied in progress \cite{Working}.

\section{Conclusions}
\label{sec:Discussions}
In summary, we have studied the weak decays of b-baryon to a tetraquark and a light baryon, decay amplitudes for various transitions have been parametrized in terms of the SU(3)-independent amplitudes. Using these results, we provide a number of relations which can be served as tests for finding out the suitable theoretical explanation for $Z_c$ states. The partial decay widths as well as branching fractions of several decay channels have also been presented in terms of the helicity amplitude technique.

At present, with limited data it is not possible to distinguish the mechanism by which the component quarks are held together to form $Z_c$ states. Since the discovery of X(3872) \cite{Belle:2003nnu}, over nearly 20 years, we still have no satisfactory landscape about the precise structures of ``XYZ'' states. Progress in clarifying this picture requires measurements of improved precision and searches for additional states. To obtain further insights of the nature of exotic charged and neutral $Z_c$ states, we suggest measurements of their production rates in the weak decays of b-baryons by the LHCb experiments in the future. In recent years, LHC has helped us find out doubly heavy baryon and pentaquark states \cite{Aaij:2017ueg,Aaij:2018gfl,LHCb:2015yax}, undoubtedly, it will provide a sustained progress in heavy baryon field. Naturally, we wish it would also provide a new milestone in the research of $Z_c$ states which will help resolve the current and longstanding puzzles in the exotic charmonium sector.

\section*{Acknowledgements}
The authors are grateful to Professor Wei Wang, Dr. Wen-Cheng Yan and Dr. Ya-Teng Zhang for inspiring and fruitful discussions and valuable comments. J.X. is supported in part by National Natural Science Foundation of China under Grant No. 12105247 and 12047545, the China Postdoctoral Science Foundation under Grant No. 2021M702957. F.H is supported in part by National Natural Science Foundation of China under grant Nos. 11735010, U2032102, 11653003, Natural Science Foundation of Shanghai under grant No. 15DZ2272100. Y.X is supported in part by National Natural Science Foundation of China under grant No. 12005294.


\begin{thebibliography}{}
\bibitem{BESIII:2013ris}
M.~Ablikim \textit{et al.} [BESIII],
Phys. Rev. Lett. \textbf{110}, 252001 (2013)
doi:10.1103/PhysRevLett.110.252001
[arXiv:1303.5949 [hep-ex]].
\bibitem{Belle:2013yex}
Z.~Q.~Liu \textit{et al.} [Belle],
Phys. Rev. Lett. \textbf{110}, 252002 (2013)
[erratum: Phys. Rev. Lett. \textbf{111}, 019901 (2013)]
doi:10.1103/PhysRevLett.110.252002
[arXiv:1304.0121 [hep-ex]].
\bibitem{Xiao:2013iha}
T.~Xiao, S.~Dobbs, A.~Tomaradze and K.~K.~Seth,
Phys. Lett. B \textbf{727}, 366-370 (2013)
doi:10.1016/j.physletb.2013.10.041
[arXiv:1304.3036 [hep-ex]].
\bibitem{BESIII:2015cld}
M.~Ablikim \textit{et al.} [BESIII],
Phys. Rev. Lett. \textbf{115}, no.11, 112003 (2015)
doi:10.1103/PhysRevLett.115.112003
[arXiv:1506.06018 [hep-ex]].
\bibitem{Wang:2013cya}
Q.~Wang, C.~Hanhart and Q.~Zhao,
Phys. Rev. Lett. \textbf{111}, no.13, 132003 (2013)
doi:10.1103/PhysRevLett.111.132003
[arXiv:1303.6355 [hep-ph]].
\bibitem{Guo:2013sya}
F.~K.~Guo, C.~Hidalgo-Duque, J.~Nieves and M.~P.~Valderrama,
Phys. Rev. D \textbf{88}, 054007 (2013)
doi:10.1103/PhysRevD.88.054007
[arXiv:1303.6608 [hep-ph]].
\bibitem{Li:2013xia}
G.~Li,
Eur. Phys. J. C \textbf{73}, no.11, 2621 (2013)
doi:10.1140/epjc/s10052-013-2621-5
[arXiv:1304.4458 [hep-ph]].
\bibitem{Zhang:2013aoa}
J.~R.~Zhang,
Phys. Rev. D \textbf{87}, no.11, 116004 (2013)
doi:10.1103/PhysRevD.87.116004
[arXiv:1304.5748 [hep-ph]].
\bibitem{Dias:2013xfa}
J.~M.~Dias, F.~S.~Navarra, M.~Nielsen and C.~M.~Zanetti,
Phys. Rev. D \textbf{88}, no.1, 016004 (2013)
doi:10.1103/PhysRevD.88.016004
[arXiv:1304.6433 [hep-ph]].
\bibitem{Braaten:2013boa}
E.~Braaten,
Phys. Rev. Lett. \textbf{111}, 162003 (2013)
doi:10.1103/PhysRevLett.111.162003
[arXiv:1305.6905 [hep-ph]].
\bibitem{Wilbring:2013cha}
E.~Wilbring, H.~W.~Hammer and U.~G.~Mei\ss{}ner,
Phys. Lett. B \textbf{726}, 326-329 (2013)
doi:10.1016/j.physletb.2013.08.059
[arXiv:1304.2882 [hep-ph]].
\bibitem{Chen:2013coa}
D.~Y.~Chen, X.~Liu and T.~Matsuki,
Phys. Rev. D \textbf{88}, no.3, 036008 (2013)
doi:10.1103/PhysRevD.88.036008
[arXiv:1304.5845 [hep-ph]].
\bibitem{Liu:2013rxa}
Y.~R.~Liu,
Phys. Rev. D \textbf{88}, 074008 (2013)
doi:10.1103/PhysRevD.88.074008
[arXiv:1304.7467 [hep-ph]].
\bibitem{Wang:2013hga}
Q.~Wang, C.~Hanhart and Q.~Zhao,
Phys. Lett. B \textbf{725}, no.1-3, 106-110 (2013)
doi:10.1016/j.physletb.2013.06.049
[arXiv:1305.1997 [hep-ph]].
\bibitem{Ali:2011ug}
A.~Ali, C.~Hambrock and W.~Wang,
Phys. Rev. D \textbf{85}, 054011 (2012)
doi:10.1103/PhysRevD.85.054011
[arXiv:1110.1333 [hep-ph]].
\bibitem{Liu:2013vfa}
X.~H.~Liu and G.~Li,
Phys. Rev. D \textbf{88}, 014013 (2013)
doi:10.1103/PhysRevD.88.014013
[arXiv:1306.1384 [hep-ph]].
\bibitem{Chen:2013wca}
D.~Y.~Chen, X.~Liu and T.~Matsuki,
Phys. Rev. Lett. \textbf{110}, no.23, 232001 (2013)
doi:10.1103/PhysRevLett.110.232001
[arXiv:1303.6842 [hep-ph]].
\bibitem{Aceti:2014uea}
F.~Aceti, M.~Bayar, E.~Oset, A.~Martinez Torres, K.~P.~Khemchandani, J.~M.~Dias, F.~S.~Navarra and M.~Nielsen,
Phys. Rev. D \textbf{90}, no.1, 016003 (2014)
doi:10.1103/PhysRevD.90.016003
[arXiv:1401.8216 [hep-ph]].
\bibitem{Wang:2013vex}
Z.~G.~Wang and T.~Huang,
Phys. Rev. D \textbf{89}, no.5, 054019 (2014)
doi:10.1103/PhysRevD.89.054019
[arXiv:1310.2422 [hep-ph]].
\bibitem{Chen:2015igx}
D.~Y.~Chen and Y.~B.~Dong,
Phys. Rev. D \textbf{93}, no.1, 014003 (2016)
doi:10.1103/PhysRevD.93.014003
[arXiv:1510.00829 [hep-ph]].
\bibitem{Huang:2015xud}
Y.~Huang, J.~He, X.~Liu, H.~F.~Zhang, J.~J.~Xie and X.~R.~Chen,
Phys. Rev. D \textbf{93}, no.3, 034022 (2016)
doi:10.1103/PhysRevD.93.034022
[arXiv:1512.00981 [hep-ph]].
\bibitem{Zhang:2020vfv}
Z.~Zhang, L.~Zheng, G.~Chen, H.~G.~Xu, D.~M.~Zhou, Y.~L.~Yan and B.~H.~Sa,
Eur. Phys. J. C \textbf{81}, no.3, 198 (2021)
doi:10.1140/epjc/s10052-021-08983-3
[arXiv:2010.10062 [hep-ph]].
\bibitem{D0:2018wyb}
V.~M.~Abazov \textit{et al.} [D0],
Phys. Rev. D \textbf{98}, no.5, 052010 (2018)
doi:10.1103/PhysRevD.98.052010
[arXiv:1807.00183 [hep-ex]].
\bibitem{LHCb:2015yax}
R.~Aaij \textit{et al.} [LHCb],
Phys. Rev. Lett. \textbf{115}, 072001 (2015)
doi:10.1103/PhysRevLett.115.072001
[arXiv:1507.03414 [hep-ex]].
\bibitem{Savage:1989ub}
  M.~J.~Savage and M.~B.~Wise,
  Phys.\ Rev.\ D {\bf 39}, 3346 (1989)
  Erratum: [Phys.\ Rev.\ D {\bf 40}, 3127 (1989)].
  doi:10.1103/PhysRevD.39.3346, 10.1103/PhysRevD.40.3127
\bibitem{Gronau:1994rj}
  M.~Gronau, O.~F.~Hernandez, D.~London and J.~L.~Rosner,
  Phys.\ Rev.\ D {\bf 50}, 4529 (1994)
  doi:10.1103/PhysRevD.50.4529
  [hep-ph/9404283].
\bibitem{He:1998rq}
  X.~G.~He,
  Eur.\ Phys.\ J.\ C {\bf 9}, 443 (1999)
  doi:10.1007/s100529900064
  [hep-ph/9810397].
\bibitem{Deshpande:2000jp}
  N.~G.~Deshpande, X.~G.~He and J.~Q.~Shi,
  Phys.\ Rev.\ D {\bf 62}, 034018 (2000)
  doi:10.1103/PhysRevD.62.034018
  [hep-ph/0002260].
\bibitem{Deshpande:1994ii}
  N.~G.~Deshpande and X.~G.~He,
  Phys.\ Rev.\ Lett.\  {\bf 75}, 1703 (1995)
  doi:10.1103/PhysRevLett.75.1703
  [hep-ph/9412393].
\bibitem{Chiang:2003pm}
  C.~W.~Chiang, M.~Gronau, Z.~Luo, J.~L.~Rosner and D.~A.~Suprun,
  Phys.\ Rev.\ D {\bf 69}, 034001 (2004)
  doi:10.1103/PhysRevD.69.034001
  [hep-ph/0307395].
\bibitem{Li:2007bh}
  Y.~Li, C.~D.~L\"u and W.~Wang,
  Phys.\ Rev.\ D {\bf 77}, 054001 (2008)
  doi:10.1103/PhysRevD.77.054001
  [arXiv:0711.0497 [hep-ph]].
\bibitem{Zhou:2015jba}
S.~H.~Zhou, Y.~B.~Wei, Q.~Qin, Y.~Li, F.~S.~Yu and C.~D.~Lu,
Phys. Rev. D \textbf{92}, no.9, 094016 (2015)
doi:10.1103/PhysRevD.92.094016
[arXiv:1509.04060 [hep-ph]].
\bibitem{Jiang:2017zwr}
H.~Y.~Jiang, F.~S.~Yu, Q.~Qin, H.~n.~Li and C.~D.~L\"u,
Chin. Phys. C \textbf{42}, no.6, 063101 (2018)
doi:10.1088/1674-1137/42/6/063101
[arXiv:1705.07335 [hep-ph]].
\bibitem{Wang:2008rk}
  W.~Wang, Y.~M.~Wang, D.~S.~Yang and C.~D.~L\"u,
  Phys.\ Rev.\ D {\bf 78}, 034011 (2008)
  doi:10.1103/PhysRevD.78.034011
  [arXiv:0801.3123 [hep-ph]].
\bibitem{Chiang:2008zb}
  C.~W.~Chiang and Y.~F.~Zhou,
  JHEP {\bf 0903}, 055 (2009)
  doi:10.1088/1126-6708/2009/03/055
  [arXiv:0809.0841 [hep-ph]].
\bibitem{Cheng:2014rfa}
  H.~Y.~Cheng, C.~W.~Chiang and A.~L.~Kuo,
  Phys.\ Rev.\ D {\bf 91}, no. 1, 014011 (2015)
  doi:10.1103/PhysRevD.91.014011
  [arXiv:1409.5026 [hep-ph]].
\bibitem{He:2014xha}
  X.~G.~He, G.~N.~Li and D.~Xu,
  Phys.\ Rev.\ D {\bf 91}, no. 1, 014029 (2015)
  doi:10.1103/PhysRevD.91.014029
  [arXiv:1410.0476 [hep-ph]].
\bibitem{He:2015fwa}
  X.~G.~He and G.~N.~Li,
  Phys.\ Lett.\ B {\bf 750}, 82 (2015)
  doi:10.1016/j.physletb.2015.08.048
  [arXiv:1501.00646 [hep-ph]].
\bibitem{He:2015fsa}
  M.~He, X.~G.~He and G.~N.~Li,
  Phys.\ Rev.\ D {\bf 92}, no. 3, 036010 (2015)
  doi:10.1103/PhysRevD.92.036010
  [arXiv:1507.07990 [hep-ph]].
\bibitem{Lu:2016ogy}
  C.~D.~L\"u, W.~Wang and F.~S.~Yu,
  Phys.\ Rev.\ D {\bf 93}, no. 5, 056008 (2016)
  doi:10.1103/PhysRevD.93.056008
  [arXiv:1601.04241 [hep-ph]].
\bibitem{Cheng:2016ejf}
  H.~Y.~Cheng, C.~W.~Chiang and A.~L.~Kuo,
  Phys.\ Rev.\ D {\bf 93}, no. 11, 114010 (2016)
  doi:10.1103/PhysRevD.93.114010
  [arXiv:1604.03761 [hep-ph]].
\bibitem{Cheng:2012xb}
  H.~Y.~Cheng and C.~W.~Chiang,
  Phys.\ Rev.\ D {\bf 86}, 014014 (2012)
  doi:10.1103/PhysRevD.86.014014
  [arXiv:1205.0580 [hep-ph]].
\bibitem{Li:2012cfa}
  H.~N.~Li, C.~D.~L\"u and F.~S.~Yu,
  Phys.\ Rev.\ D {\bf 86}, 036012 (2012)
  doi:10.1103/PhysRevD.86.036012
  [arXiv:1203.3120 [hep-ph]].
\bibitem{Qin:2013tje}
Q.~Qin, H.~n.~Li, C.~D.~L\"u and F.~S.~Yu,
Phys. Rev. D \textbf{89}, no.5, 054006 (2014)
doi:10.1103/PhysRevD.89.054006
[arXiv:1305.7021 [hep-ph]].
\bibitem{Wang:2017azm}
W.~Wang, Z.~P.~Xing and J.~Xu,
Eur. Phys. J. C \textbf{77}, no.11, 800 (2017)
doi:10.1140/epjc/s10052-017-5363-y
[arXiv:1707.06570 [hep-ph]].
\bibitem{Shi:2017dto}
Y.~J.~Shi, W.~Wang, Y.~Xing and J.~Xu,
Eur. Phys. J. C \textbf{78}, no.1, 56 (2018)
doi:10.1140/epjc/s10052-018-5532-7
[arXiv:1712.03830 [hep-ph]].
\bibitem{Wang:2018utj}
W.~Wang and J.~Xu,
Phys. Rev. D \textbf{97}, no.9, 093007 (2018)
doi:10.1103/PhysRevD.97.093007
[arXiv:1803.01476 [hep-ph]].
\bibitem{Li:2021rfj}
D.~M.~Li, X.~R.~Zhang, Y.~Xing and J.~Xu,
Eur. Phys. J. Plus \textbf{136}, no.7, 772 (2021)
doi:10.1140/epjp/s13360-021-01757-6
[arXiv:2101.12574 [hep-ph]].
\bibitem{Zhang:2018llc}
Q.~A.~Zhang,
Eur. Phys. J. C \textbf{78}, no.12, 1024 (2018)
doi:10.1140/epjc/s10052-018-6481-x
[arXiv:1811.02199 [hep-ph]].
\bibitem{Huang:2021jxt}
F.~Huang, J.~Xu and X.~R.~Zhang,
Eur. Phys. J. C \textbf{81}, no.11, 976 (2021)
doi:10.1140/epjc/s10052-021-09729-x
[arXiv:2107.13958 [hep-ph]].
\bibitem{Chen:2022asf}
H.~X.~Chen, W.~Chen, X.~Liu, Y.~R.~Liu and S.~L.~Zhu,
[arXiv:2204.02649 [hep-ph]].
\bibitem{He:2006ud}
X.~G.~He, T.~Li, X.~Q.~Li and Y.~M.~Wang,
Phys. Rev. D \textbf{74}, 034026 (2006)
doi:10.1103/PhysRevD.74.034026
[arXiv:hep-ph/0606025 [hep-ph]].
\bibitem{Xing:2022uqu}
Z.~P.~Xing, F.~Huang and W.~Wang,
[arXiv:2203.13524 [hep-ph]].
\bibitem{Rui:2022sdc}
Z.~Rui, C.~Q.~Zhang, J.~M.~Li and M.~K.~Jia,
[arXiv:2206.04501 [hep-ph]].
\bibitem{Buchalla:1995vs}
G.~Buchalla, A.~J.~Buras and M.~E.~Lautenbacher,
Rev. Mod. Phys. \textbf{68}, 1125-1144 (1996)
doi:10.1103/RevModPhys.68.1125
[arXiv:hep-ph/9512380 [hep-ph]].
\bibitem{Wang:2008sm}
Y.~m.~Wang, Y.~Li and C.~D.~Lu,
Eur. Phys. J. C \textbf{59}, 861-882 (2009)
doi:10.1140/epjc/s10052-008-0846-5
[arXiv:0804.0648 [hep-ph]].







\bibitem{Agaev:2017tzv}
S.~S.~Agaev, K.~Azizi and H.~Sundu,
Phys. Rev. D \textbf{96}, no.3, 034026 (2017)
doi:10.1103/PhysRevD.96.034026
[arXiv:1706.01216 [hep-ph]].
\bibitem{Mott:2011cx}
L.~Mott and W.~Roberts,
Int. J. Mod. Phys. A \textbf{27}, 1250016 (2012)
doi:10.1142/S0217751X12500169
[arXiv:1108.6129 [nucl-th]].
\bibitem{Wang:2015ndk}
Y.~M.~Wang and Y.~L.~Shen,
JHEP \textbf{02}, 179 (2016)
doi:10.1007/JHEP02(2016)179
[arXiv:1511.09036 [hep-ph]].
\bibitem{Wang:2009hra}
Y.~M.~Wang, Y.~L.~Shen and C.~D.~Lu,
Phys. Rev. D \textbf{80}, 074012 (2009)
doi:10.1103/PhysRevD.80.074012
[arXiv:0907.4008 [hep-ph]].
\bibitem{Hsiao:2015cda}
Y.~K.~Hsiao, P.~Y.~Lin, C.~C.~Lih and C.~Q.~Geng,
Phys. Rev. D \textbf{92}, 114013 (2015)
doi:10.1103/PhysRevD.92.114013
[arXiv:1509.05603 [hep-ph]].
\bibitem{Workman:2022ynf}
R.~L.~Workman \textit{et al.} [Particle Data Group],
PTEP \textbf{2022}, 083C01 (2022)
doi:10.1093/ptep/ptac097
\bibitem{Lu:2000em}
C.~D.~Lu, K.~Ukai and M.~Z.~Yang,
Phys. Rev. D \textbf{63}, 074009 (2001)
doi:10.1103/PhysRevD.63.074009
[arXiv:hep-ph/0004213 [hep-ph]].
\bibitem{Chou:2001bn}
C.~H.~Chou, H.~H.~Shih, S.~C.~Lee and H.~n.~Li,
Phys. Rev. D \textbf{65}, 074030 (2002)
doi:10.1103/PhysRevD.65.074030
[arXiv:hep-ph/0112145 [hep-ph]].
\bibitem{Beneke:1999br}
M.~Beneke, G.~Buchalla, M.~Neubert and C.~T.~Sachrajda,
Phys. Rev. Lett. \textbf{83}, 1914-1917 (1999)
doi:10.1103/PhysRevLett.83.1914
[arXiv:hep-ph/9905312 [hep-ph]].
\bibitem{Bauer:2000yr}
C.~W.~Bauer, S.~Fleming, D.~Pirjol and I.~W.~Stewart,
Phys. Rev. D \textbf{63}, 114020 (2001)
doi:10.1103/PhysRevD.63.114020
[arXiv:hep-ph/0011336 [hep-ph]].
\bibitem{Bauer:2001yt}
C.~W.~Bauer, D.~Pirjol and I.~W.~Stewart,
Phys. Rev. D \textbf{65}, 054022 (2002)
doi:10.1103/PhysRevD.65.054022
[arXiv:hep-ph/0109045 [hep-ph]].
\bibitem{Working}
F.~Huang, Y.~Xing and J.~Xu,
\textit{work in progress.}
\bibitem{Belle:2003nnu}
S.~K.~Choi \textit{et al.} [Belle],
Phys. Rev. Lett. \textbf{91}, 262001 (2003)
doi:10.1103/PhysRevLett.91.262001
[arXiv:hep-ex/0309032 [hep-ex]].
\bibitem{Aaij:2017ueg}
R.~Aaij \textit{et al.} [LHCb],
Phys. Rev. Lett. \textbf{119}, no.11, 112001 (2017)
doi:10.1103/PhysRevLett.119.112001
[arXiv:1707.01621 [hep-ex]].
\bibitem{Aaij:2018gfl}
R.~Aaij \textit{et al.} [LHCb],
Phys. Rev. Lett. \textbf{121}, no.16, 162002 (2018)
doi:10.1103/PhysRevLett.121.162002
[arXiv:1807.01919 [hep-ex]].







\end{thebibliography}
\end{document}